\documentclass[10pt,letterpaper]{article}

\usepackage[top=0.85in,left=2.75in,footskip=0.75in]{geometry}

\usepackage{amsmath,amssymb}
\usepackage{changepage}
\usepackage[utf8x]{inputenc}
\usepackage{textcomp,marvosym}
\usepackage{cite}
\usepackage{nameref,hyperref}
\usepackage[right]{lineno}
\usepackage{microtype}
\usepackage{float}
\DisableLigatures[f]{encoding = *, family = * }
\usepackage[table]{xcolor}
\usepackage{array}
\newcolumntype{+}{!{\vrule width 2pt}}
\newlength\savedwidth

\raggedright
\usepackage[aboveskip=1pt,labelfont=bf,labelsep=period,justification=raggedright,singlelinecheck=off]{caption}

\makeatletter
\renewcommand{\@biblabel}[1]{\quad#1.}
\makeatother

\usepackage{lastpage,fancyhdr,graphicx}
\usepackage{epstopdf}
\fancyheadoffset[L]{2.25in}
\fancyfootoffset[L]{2.25in}
\newif\ifshowfigures
\newcommand{\figonoff}[1]{\ifshowfigures{#1} \fi}

\showfigurestrue %% Show figures

\renewcommand{\d}{\mathrm{d}}
\newcommand{\e}{\mathrm{e}}
\usepackage{graphics}

\begin{document}
\vspace*{0.2in}

\begin{flushleft}
{\Large \textbf\newline{Modeling the evolution of COVID-19 via
    compartmental and particle-based approaches: application to the
    Cyprus case} }
\newline
\\
Constantia Alexandrou\textsuperscript{1,2}\textsuperscript{\P},
Vangelis Harmandaris\textsuperscript{2,3}\textsuperscript{\P},
Anastasios Irakleous\textsuperscript{1,2}\textsuperscript{\P},
Giannis Koutsou\textsuperscript{2*}\textsuperscript{\P},
Nikos Savva\textsuperscript{2}\textsuperscript{\P}
\\
\bigskip
\textbf{1} Department of Physics,  University of Cyprus, P/O/ Box 20537, 1678 Nicosia, Cyprus 
\\
\textbf{2} Computation-based Science and Technology Research Center, The Cyprus Institute, 20 Constantinou Kavafi Str., Nicosia 2121, Cyprus
\\
\textbf{3} Department of Mathematics and Applied Mathematics, University of Crete, \& Institute of Applied and Computational Mathematics, FORTH, GR-71110 Heraklion, Greece
\bigskip

\textsuperscript{\P} All authors contributed equally to this work and they thus appear in alphabetical order.

* Corresponding author G. Koutsou, email:g.koutsou@cyi.ac.cy

\end{flushleft}
\section*{Abstract}
We present two different approaches for modeling the spread of the
COVID-19 pandemic. Both approaches are based on the population classes
susceptible, exposed, infectious, quarantined, and recovered and allow
for an arbitrary number of subgroups with different infection rates
and different levels of testing. The first model is derived from a set
of ordinary differential equations that incorporates the rates at
which population transitions take place among classes. The other is a
particle model, which is a specific case of crowd simulation model, in
which the disease is transmitted through particle collisions and
infection rates are varied by adjusting the particle velocities. The
parameters of these two models are tuned using information on COVID-19
from the literature and country-specific data, including the effect of
restrictions as they were imposed and lifted. We demonstrate the
applicability of both models using data from Cyprus, for which we find
that both models yield very similar results, giving confidence in the
predictions.

\section{Introduction}
The COVID-19 pandemic is a new disease and there is as yet not enough
understanding on its future evolution. Since medical interventions,
such as vaccines or antiviral treatments, are not available yet,
non-medical interventions are being implemented to contain the
disease. In a number of countries, including Cyprus, the imposed
restrictions have helped slow down the spread of the disease. Cyprus,
being an island country, managed to limit the spread of the disease,
by imposing restrictions on air travel and shutting down large parts
of the economy, ultimately achieving 2.2 deaths per 100,000
population, a death rate comparable to that of Greece and
Malta~\cite{JHC,owidcoronavirus}. However, as restrictions are being
lifted, it is important to know how the disease in each country will
evolve. Reliable predictions will help policy-makers to formulate
appropriate intervention strategies, while taking into account also
economic and social factors. Mathematical models and numerical
simulation can be used as a decision support tool to assist
policy-makers, by forecasting the spread of the disease as a function
of the lifting of restrictions as well as on the level of testing and
contact
tracing~\cite{Ferguson2020,Wang2020,Kupferschmidt2020,Giordano2020,Gatto2020,Ferguson2020b,
  PPR:PPR113604, DAS2021110595,Keskinocak2020,Keskinocak2020b}. Since the
epidemic spreading is a complex process depending to a large extent
both on the behavior of the virus~\cite{Wang2020, Kupferschmidt2020}
and human interactions~\cite{Giordano2020,Gatto2020}, the purpose of
this work is to provide such predictions for a number of
\textit{scenarios} using two models. While these models are applicable
to any country, they are calibrated for the specific case of Cyprus
for which such forecasting is not available.

Since the outbreak of COVID-19, a number of mathematical
epidemiological models have been used to predict the spread of the
disease in a number of countries~(e.g. Refs.~\cite{Ferguson2020,
  Ferguson2020b, Gatto2020, PPR:PPR113604, DAS2021110595}) and regions~(e.g.,
Refs.~\cite{Keskinocak2020,Keskinocak2020b}).  Typically, the
predictions are made within a single type of model.  In this work, we
use two models, each relying on different methodologies, to
cross-check predictions.
\begin{enumerate}
\item \textit{Compartmental models based on ordinary differential
  equations (ODEs)}. These models describe how portions of the
  population transition among classes or compartments via ODEs. For
  example, the classical SIR model describes the evolution of three
  compartments, the susceptible, infected, and removed~\cite{SIR1927},
  with two parameters that model the rates at which the population
  transfers from one compartment to the other. In the present study,
  our objective is to derive a model that is suitable for countries
  like Cyprus, where data typically used for modeling COVID-19, such
  as of hospitalizations, intubation, and deaths, are too small for a
  meaningful data-driven analysis. We therefore use a time-dependent
  infection rate and detection rate, with which we are able to capture
  the reported cases in Cyprus without needing to revert to models
  with a large number of parameters relying on data that are scarce or
  not available.
\item \textit{Particle model}. Such models belong to the broader class
  of crowd simulation/agent-based models, which track the evolution of
  the interactions of agents in time and space. Here, particle
  dynamics and interactions are used as a proxy of human social
  interactions, allowing the transmission of the disease through
  particle collisions, naturally introducing an element of randomness
  in the system. This approach is reminiscent of the equivalent of
  network models and stochastic-branching processes, which is often
  used to model disease outbreaks.  Such models are computationally
  demanding and an efficient code will be employed in order to
  simulate the whole population of Cyprus residing in cities, where
  most of the COVID-19 cases are registered.
\end{enumerate}
The main goal of the study is to examine the forecasting potential of
the above models for the short-term evolution of COVID-19, under
various conditions related to imposing or lifting measures.  We
demonstrate that with a relatively small set of parameters both models
describe quite accurately the existing data and yield very similar
predictions for the evolution of the disease when the same assumptions
are made for the measures imposed and lifted. More importantly, we
examine whether such models predict the number of positive COVID-19
cases for small countries like Cyprus assuming a number of different
scenarios for the evolution of the infection rate, as the economy
reopens and in relation to the ability of the health care services to
detect and isolate future cases through contact tracing and aggressive
testing. One of the key highlights of this study is that, although
these models are based on different approaches, they both yield
consistent predictions within their corresponding uncertainties. This
leads to confidence on the model forecasts on one hand, but also
allows us to attribute different errors to the predictions, since the
particle model exhibits stochastic errors, related to how the pandemic
was seeded within the population, as well as uncertainties in the
parameters, such as the velocities of the particles, while the
compartmental model exhibits modeling uncertainty, based on the
uncertainties of the fitted parameters. This study also demonstrates
the ability of these models to capture features of the pandemic, such
as the infection rate, when relatively small numbers of cases are
involved and for which statistical models yield results with very
large uncertainties. Their reliability, on the other hand, is also
subject to changes in the way the population collectively behaves
during the forecasting period.

The structure of the paper is as follows: In Sec.~\ref{sec:models} we
introduce the two models and describe how we tune the parameters,
adjusting them to the specific case of Cyprus. In
Sec.~\ref{sec:results} we discuss the results of the models and in
Sec.~\ref{sec:conclusions} we give our conclusions and future plans.

\section{Description of the models}\label{sec:models}
We describe here the two models and how their parameters are
tuned. There is presently a plethora of models of varying degrees of
complexity and sophistication, depending on the wealth of data
available. For example, for larger populations, where naturally the
death toll is higher, modeling approaches may be effectively used to
capture the evolution of the number of deceased~\cite{CARLETTI2020100034,FANELLI2020109761,Dehningeabb9789}.
For smaller
countries, however, like Cyprus with a small number of deaths, the
applicability of e.g. statistical models that focus on extracting
multitude of parameters from the recorded deaths~\cite{Ferguson2020}
may be questionable. Here we propose an approach that uses a minimal
set of fitting parameters. To model COVID-19 evolution one needs to
make certain assumptions.  In our modeling we use some common features
that are also being applied in a number of other studies that are
currently being applied to
COVID-19~\cite{Ferguson2020, Ferguson2020b, Gatto2020, PPR:PPR113604, DAS2021110595}.
Namely, we consider that

\begin{description}
\item[i] The disease consists of two phases: (a) an exposed phase
  during which an individual contracts the disease but has no symptoms
  and does not transmit the disease to others; (b) an infectious
  phase, during which individuals transmit the disease to susceptible
  members of the population.
\item[ii] A portion of the exposed individuals who ultimately become
  infectious is detected. We take this portion to be a function of
  time depending on the level of screening and testing being
  performed.
\item[iii] The detected cases are assumed to be immediately
  quarantined, so that the spread of the virus is attributed solely to
  the undetected cases.
\item[iv] The population is divided in five classes; the susceptible,
  exposed, quarantined, infectious, and removed; i.e. an SEIQR type of
  model.
\item[v] The removed class includes individuals who (a) recover or (b)
  die due to the disease and we do not differentiate among these
  subclasses. Deaths can be modeled by inputing the death rate, which
  also depends on the capacity of healthcare systems to accommodate
  the needs of those who become critically ill. While in the future
  such an analysis may be possible, at the time of writing, there are
  not enough data available for the case of Cyprus to reliably extract
  a death rate. In addition, the quarantined class is modeled in the
  same way as the removed class, but tracked independently as a
  function of time
\item[vi] The average duration of the exposed latent period is constant.
\item[vii] The average duration of the infectious period is constant.
\item[viii] A key feature of our modeling is that we take the rate of
  infection/transmission to be a function of time, reflecting the
  measures implemented by the government in regards to social
  distancing, limitations of travel, and suspending parts of the
  economy.
\end{description}

For the infectious period, we rely on recent data that estimate it to
be 7-12
days.~\cite{ecdpc}.
We therefore fix the infectious period ($\tau_i$) to 10~days and the
exposed period ($\tau_e$) to 2~days.

These modeling features are kept the same for both our models. Further
extensions of the models could include, for example, assigning a
probability of the quarantined population to infect, a delay between
the time an infectious individual is quarantined, or defining
additional groups within the classes which would however introduce
more parameters. While such extensions of the models may be considered
in the future as more data become available, we opt to evaluate here
our models with the least possible parameters, which as will be seen
in Sec.~\ref{sec:results} are sufficient to describe the current data.

\subsection{Compartmental model}\label{sec:comp}
Extending the original SIR model of Kermack and McKendrick
\cite{SIR1927} to include an Exposed and a Quarantined class, the
following coupled system of ODEs arises, which is in alignment with
the assumptions introduced earlier:
\begin{subequations}\label{eq:seircont}
\arraycolsep=1.4pt
\begin{eqnarray}
\frac{\d S}{\d t}&=&-\beta \dfrac{SI}{N},\\
\frac{\d E}{\d t}&=& \beta \dfrac{SI}{N}-\dfrac{E}{\tau_e},\\
\frac{\d I}{\d t}&=& r \dfrac{E}{\tau_e} - \dfrac{I}{\tau_i},\\
\frac{\d Q}{\d t}&=& (1-r) \dfrac{E}{\tau_e} - \dfrac{Q}{\tau_i},\\
\frac{\d R}{\d t}&=&\dfrac{I+Q}{\tau_i},
\end{eqnarray}
\end{subequations}
where $S(t)$, $E(t)$, $I(t)$, $Q(t)$, and $R(t)$ capture,
respectively, the evolution of the susceptible, exposed, infected,
quarantined, and removed classes of the population as a function of
time $t$. We will refer to this extended model as the SEIQR model.
The parameter $\beta>0$ corresponds to the infection rate in inverse
time units, which is a measure of the average number of contacts an
infective individual makes in a wholly susceptible population that may
lead to an infection per unit time, $\tau_{i}>0$ and $\tau_{e}>0$ are
the average times an individual remains in the infective and exposed
classes respectively. The portion of undetected cases who later infect
others is given by $r$ and $N$ is the total population (assuming no
vital dynamics, namely births or deaths due to other causes).

Central to the standard SIR and SEIQR models is the modeling of the
rates at which individuals transition between population classes.
Both the SIR and SEIQR models assume that the times for which
individuals remain in the exposed and infectious states are
exponentially distributed random variables~\cite{Hethcote1980}, which
implies that the chance of an individual moving out of the exposed and
recovered classes is independent of the time they entered the
particular class. This leads to dispersed timescales, which manifests
itself, for example, in unrealistically long recovery times for the
number of individuals that got infected towards the end of an epidemic
and in overoptimistic predictions of the levels of control required to
contain the epidemic~\cite{Wearing2005}. A more general approach
assigns arbitrary probability distributions for the recovery and
latent times. This yields a system of integral--differential equations
of the form

\vspace*{-\baselineskip}
\begin{subequations}\label{eq:int}
\begin{align}
  &E(t)=E(0)P_E(t) + \displaystyle\int_0^t \beta \frac{S(\tau)I(\tau)}{N}P_E(t-\tau)\,\d \tau,\\
  &I(t) = I(0)P_I(t)+ r \displaystyle\int_0^t \left(\beta \frac{S(\tau)I(\tau)}{N}- \frac{\d }{\d \tau} E(\tau)\right)P_I(t-\tau)\,\d \tau,\\
  &Q(t) = Q(0)P_I(t) +(1-r) \displaystyle\int_0^t \left(\beta \frac{S(\tau)I(\tau)}{N}-\frac{\d }{\d \tau} E(\tau)\right)P_I(t-\tau)\,\d \tau,\\  
  &R(t) =(Q(0)+I(0))\left( 1-P_I(t) \right)- \displaystyle\int_0^t\! \left(\beta \frac{S(\tau)I(\tau)}{N}-\frac{\d E(\tau)}{\d \tau}\! \right)\! \left(1-P_I(t-\tau)\right)\,\d \tau,\\
  &S(t)+E(t)+I(t)+R(t)=N,
\end{align}
\end{subequations}
where $P_{I}(t)$ and $P_{E}(t)$ are non-increasing functions that
correspond to the probabilities of remaining infectious or quarantined
and exposed $t$ units after becoming infectious or quarantined and
exposed, respectively with $P_{I}(0)=P_E(0)=1$ (see
\cite{Hethcote1980} for a related model). In each of
Eqs.~(\ref{eq:int}a--c), the first terms correspond to the respective
initial populations in a class that remain in the same class after $t$
time units, whereas the second terms correspond to the sum of
individuals who become members of a class within the time interval
$[0,t]$. Eq.~(\ref{eq:int}d) captures the transfer of the infectious
and quarantined classes to the removed classes, whereas combining
Eqs.~(\ref{eq:int}a--e) yields Eq.~(\ref{eq:seircont}a).

As noted in earlier works (e.g. Ref.~\cite{Hethcote1980}), letting
$P_E(t)=\e^{-t/\tau_e}$ and $P_I(t)=\e^{-t/\tau_i}$ reduces
Eqs.~(\ref{eq:int}) to Eqs.~(\ref{eq:seircont}). Although it has been
argued that more general multi-stage gamma-distributed latent and
infectious periods may be more appropriate, see
e.g. \cite{Feng2000,Lloyd2001,Wearing2005}, precise knowledge of
$P_E(t)$ and $P_I(t)$ is neither known nor expected to have an
appreciable qualitative effect. Here we chose the latent and
infectious periods to be of fixed length as means to alleviate the
aforementioned issues if these are exponentially distributed, see also
Ref.~\cite{Zhang2008}. By doing so, we manage to preserve the
simplicity of the model, allowing us also to obtain a discrete set of
equations. Hence $P_E(t)$ and $P_I(t)$ are assumed to be of the form
\begin{equation}\stepcounter{equation}
P_E(t)=\begin{cases}
1,& 0\le t < \tau_e\\
0, & t\ge \tau_e
\end{cases},\qquad
P_I(t)=\begin{cases}
1,& 0\le t < \tau_i\\
0, & t\ge \tau_i
\end{cases}.\tag{\theequation a,b}
\end{equation}
Considering this time-dependence, we are able to deduce the following
system of delay differential equations
\begin{subequations}\label{eq:seir}
\begin{align}
\frac{\d S}{\d t} & =-X(t),\\
\frac{\d E}{\d t} & = X(t)-X(t-\tau_e),\\
\frac{\d I}{\d t} & = r X(t-\tau_e) - r X(t-\tau_i-\tau_e)-I_0\delta(t-\tau_i),\\
\frac{\d Q}{\d t} & = (1-r) X(t-\tau_e) - (1-r) X(t-\tau_i-\tau_e)-Q_0\delta(t-\tau_i)\\
\frac{\d R}{\d t} & = X(t-\tau_i-\tau_e)+I_0\delta(t-\tau_i) + Q_0\delta(t-\tau_i),
\end{align}
where, for $t>0$,
\begin{equation}
X(t) = \beta\dfrac{S(t)I(t)}{N}
\end{equation}
\end{subequations}
when $t>0$ and $\delta(t)$ being the Dirac delta function.

The system of Eqs.~(\ref{eq:seir}) is further extended in the present
study by incorporating the features given by items ii), iii) and vi)
above, achieved by introducing a time-dependent function
$r(t)$ which corresponds to the portion of undetected cases.

Furthermore, rather than solving the delay differential equations that
may require specialized techniques, we opt to convert the system of
Eqs.~(\ref{eq:seir}) to a discrete difference equation, using a time
step of 1, so that they become:
\begin{subequations}\label{eq:discrete_system}
\begin{align}
S(t+1)&=S(t) - X(t),\\
E(t+1)&=E(t) + X(t)-X(t-\tau_e),\\
I(t+1)&=I(t)+ r(t-\tau_e)X(t-\tau_e) - r(t-\tau_i-\tau_e)X(t-\tau_i-\tau_e) - I_0\delta_{t,\tau_i},\\
Q(t+1)&=Q(t) + (1-r(t-\tau_e)) X(t-\tau_e) \nonumber\\
&-(1-r(t-\tau_i-\tau_e))X(t-\tau_e-\tau_i) - Q_0\delta_{t,\tau_i}\\ 
R(t+1)&=R(t)+ X(t-\tau_i-\tau_e)+(Q_0+I_0)\delta_{t,\tau_i},
\end{align}
\end{subequations}
where $t$ represents discrete time in days and $\delta$ is the
Kronecker delta. The choice of the form of $r(t)$ will be discussed in
Sec.~\ref{sec:results}. The functional forms of
Eqs.~(\ref{eq:discrete_system}) are to be solved with the appropriate
initial conditions. We have confirmed numerically that such an
approach does not compromise the overall quantitative agreement with
the solutions to the original system of delay differential
equations. Note that the equations for $E(t)$, $Q(t)$ and $R(t)$ may
be decoupled from the system, as they can be fully specified
independently once the evolution of $X(t)$, $I(t)$, and $S(t)$ is
determined. Hence, it suffices to keep track of the movement of
individuals across the susceptible and infective classes as well as
the number of exposed individuals at each day. Initially, we take
$I(0)=r(0)c_0/(1-r(0))$, the actual undetected cases who are assumed
to be responsible for initiating the epidemic outbreak as derived from
the initial confirmed cases $c_0$. The exposed individuals at $t=0$
are those who become infectious at $t=\tau_e$ and hence
$X(0)=(c_2-c_1)/(1-r(0))$, where $c_1$ and $c_2$ are the confirmed
cases on days 1 and 2, respectively. Likewise, the exposed individuals
one day before the first cases are confirmed at $t=0$, will be those
who will become infected when $t=1$, so that we may take
$X(-1)=(c_1-c_0)/(1-r(0))$, assuming for simplicity that
$r(0)=r(-1)$. This allows us to determine the susceptibles at $t=0$ as
$S(0)=N-c_2/(1-r(0))$. Summarizing, we consider $X(t)$ of the form
\begin{equation}
X(t)=\begin{cases}
\dfrac{c_1-c_0}{1-r(0)},&t=-1\\[1.5em]
\dfrac{c_2-c_1}{1-r(0)},&t=0\\[1.5em]
\dfrac{\beta(t) S(t)I(t)}{N},&t>0
\end{cases},
\end{equation}
with $\beta(t)$ being time-dependent to reflect governmental measures
imposed or lifted. The fitting process is facilitated by the fact that
the government imposes or relaxes measures in $M$ stages and at given
times $t_1,\ t_2,\ \ldots,\ t_M$. In order to use the fewest possible
parameters, we take $\beta(t)$ to be of the form
\begin{equation}
  \beta(t) = \frac{1}{2}\left[b_{M} + b_0 + \sum_{j=1}^M \left(b_j-b_{j-1}\right) \tanh(m_j(t-t_j))\right],
  \label{eq:beta_t}
\end{equation}
where $b_{0}$ and $b_{M}$ being, respectively, the initial and final
transmission rates ($\lim_{t\to\infty}\beta(t) =b_M$), $b_j$,
$j=1,\ \ldots,\ M-1$ correspond to intermediate transmission rates,
and $\tanh(x) = \frac{1 - e^{-2x}}{1 + e^{-2x}}$ is the hyperbolic
tangent function. This choice for modeling the time dependence of the
infection rate is very flexible, allowing us to cover a broad range of
functional time dependence of $\beta(t)$.  Namely, the parameters
$m_j$ control how smoothly $b_{j-1}$ transitions to $b_j$ so that
$1/m_j$ gives an order of magnitude of how long this transition lasts
(as $m_j\rightarrow \infty$ this transition becomes step-like).

The model parameters are obtained by fitting the data for confirmed
cases $c_t$, to our model confirmed cases, given by:
\begin{equation}
  C(t) = c_0 +\sum_{\tau=\tau_e-1}^t (1-r(\tau-\tau_e)) X(\tau-\tau_e)
  \label{eq:confirmed_t}
\end{equation}
through a least-squares fit. To fit the initial stages until July 2020
of the Cyprus case for which this study was undertaken, i.e. during
the first lock-down and the gradual lifting of measures following it,
we consider a two-stage process ($M=2$) based on the actual
announcements by the government, with measures enforced on March
24\textsuperscript{th}, 2020 ($t_1=15$) and almost fully lifted on May
21\textsuperscript{st}, 2020 ($t_2=73$). This leads to a five
parameter fit for $b_0$, $b_1$, $b_2$, $m_1$, and $m_2$. If we were to
consider a more extended period of time, then the method can be
automated to enable the determination of $M$ and the times $t_j$ by
finding the inflection points of $c_t$, i.e. finding the times at
which the second derivative changes sign. This is done by first
smoothing out $c_t$ with a smoothing spline. Furthermore, for more
extended periods of time, $m_1$ and $m_2$ can be set to those obtained
by fitting the first lock-down phase, while $m_j$ for $j>2$ are kept
fixed leaving us with an $M+1$-parameter fit, namely to determine the
parameters $b_j$, $j=0, ..., M$. Therefore, fitting the initial period
is still relevant if one wants to use the models for a long period
during the pandemic.

The reported as recovered would require the introduction of yet
another timescale, since according to the protocols followed,
individuals are considered to have recovered after they test negative
twice within a period of 24 hours and only after all symptoms are
resolved, which leads to a median recovery time of 23
days~\cite{cygov}.  The portion of undetected $r(t)$ evolves in a
prescribed manner that captures how aggressively testing is performed,
and is elaborated in Sec.~\ref{sec:results}.

Below we present plausible scenarios for the evolution of COVID-19 in
Cyprus based on the knowledge of measures as of July 2020. We stress
that additional scenarios can be analyzed to reflect new different
circumstances as they evolve e.g. an increase of cases from incoming
people, etc. The strengths of the models is that they can be adjusted
to new measures and human behavior by adjusting the reproduction
number. Indeed, the form of the infection rate we have chosen allows
our models to capture such changes, as also demonstrated in the
addendum in which we include an example of how we can model data that
have become available after submission of this work. We should note
however, that interpreting and distinguishing between the different
factors that contribute to a given change in the reproduction number
is non-trivial and beyond the scope of this work.

\subsubsection*{Forecasting for various scenarios}
Beyond the available data, which are fitted to determine the five
parameters of the model, we forecast the evolution of the epidemic
based on different scenarios on how the infection rate $\beta(t)$ will
evolve. In one scenario, we choose to divide the three major classes
$X(t)$, $S(t)$ and $I(t)$ into two groups each, namely $X_{k}(t)$,
$S_{k}(t)$ and $I_{k}(t)$ with $k=1,2$ that can be used to model
situations where a portion of the population, e.g. people above 65
years old and/or with preexisting conditions, is considered vulnerable
to the disease and continues to observe strict social distancing
measures compared to the rest who return to work. Hence, initially and
up to some time $t=t_*$ the population is assumed to behave uniformly
to reflect the lockdown imposed to the whole population. For $t>t_*$,
we differentiate into subclasses as follows. We prescribe the
$2\times2$ contact matrix with constant entries $\beta_{j,k}$, which
denotes the average number of infectious contacts made per day by an
individual in group $j$ with an individual in group $k$. Since the
total number of contacts between group $j$ to $k$ must equal the
number of contacts from group $k$ to group $j$, we must have $N_j
\beta_{j,k}=N_k\beta_{k,j}$, where $N_1$ and $N_2$ are the
corresponding populations of each group with $N=N_1+N_2$.

Therefore, the exposed, infectious and susceptible for groups $1$ and
$2$ evolve according to
\begin{subequations}\label{eq:discrete_system2}
\arraycolsep=1.4pt
\begin{eqnarray}
X_k(t)&=&\left(\dfrac{\beta_{k,1}I_1(t)}{N_1}+\dfrac{\beta_{k,2}I_2(t)}{N_2}\right)S_k(t)\\
S_{k}(t+1)&=&S_k(t) - X_k(t),\\
I_k(t+1)&=&I_k(t)+ r_k(t-\tau_e)X_k(t-\tau_e) - r_k(t-\tau_i-\tau_e)X_k(t-\tau_i-\tau_e),
\end{eqnarray}
\end{subequations}
for $t>t_*$ and $k=1$ and $2$. For $t\le t_*$ the time histories of
$X_k$, $S_k$ and $I_k$ correspond to scalings of $X(t)$, $S(t)$ and
$I(t)$ by $N_k/N$, and we also let $r_k(t)=r(t)$. Hence, the confirmed cases for
each of the two groups, $C_k$ ($k=1$ and $2$), is given at time
$t>t_*$ by
\begin{equation}
  C_k(t) = C(t_*)\frac{N_k}{N} + \sum_{\tau=t_*+\tau_e-1}^t\left( 1-r_k(\tau-\tau_e) \right)X_k(\tau-\tau_e)
  \label{eq:two pop}
\end{equation}
Whether the two groups evolve differently depends crucially on
$r_{1}(t)$ and $r_2(t)$ and the contact matrix $\beta_{j,k}$. For
instance, if we set $r_k(t)=r(t)$ and $\beta_{k,j}=\beta N_j /N$, the
collective evolution of the two groups is not distinguishable from a
simulation using Eqs.~(\ref{eq:discrete_system}) with the two groups
mixed in a single group.

If detailed data on the decomposition of the confirmed cases are
available, this two population model, and its generalization to
multiple populations, can be used to fit and obtain the entries of the
contact matrix as fit parameters. This would allow capturing, for
example, heterogeneity between subgroups of the population as regards
their infectivity and susceptibility. However, in the case of Cyprus
such data are not available and, thus, we fit to a single population
model, and limit to using the two population model in the forecasting,
to qualitatively assess the effect of different measures being imposed
to different subgroups.

\subsection{Particle model}
\label{sec:particle}
Within the particle model, disease transmission is modeled by elastic
collisions of two-dimensional hard discs. The number of particles per
unit time scattered in any direction from a given disc is
\begin{equation}
  N_{\rm coll.}=D n v_0,
  \label{eq:collisions}
\end{equation}
where $n$ is the number of discs per unit area with radius $D/2$ and
$v_0$ is their velocity. We take $D=4$~m so that individuals with
distance greater than 2~m can not infect others. This fixes the length
unit of the system. Then the basic reproduction number is given by
\begin{equation}
\mathcal{R}_0=N_{\rm coll.} p \tau_i,
\label{eq:particle R0}
\end{equation}
where $p$ is the probability that an infectious individual transmits
the disease upon collision with another individual. We take $\tau_i$
to be the same average time a single individual is infectious as for
the extended SEIQR model.

The work-flow implemented includes using the
DynamO~\cite{doi:10.1002/jcc.21915} particle simulator within our own
post-processing scripts to generate a list of elastic particle
collisions. This list is then parsed with a given set of parameters,
namely initial number of infected, exposed, and quarantined, infection
probability for each population subgroup, detection rate as a function
of time, and velocities as a function of time. At regular time-steps
the total number of susceptible, exposed, infectious, quarantined, and
recovered are registered. As in the case of the extended SEIQR model,
an exposed individual transitions either to infectious or quarantined
based on a time-dependent undetected ratio $r(t)$. Quarantined
individuals do not infect and their time evolution is taken to model
the reported numbers by the government.  Our work-flow, which includes
post-processing of the output from DynamO, analyzing the collisions
list, and processing and visualizing the data is available
online~\cite{pSEIQR}.

\begin{figure}[!h]
  \figonoff{\includegraphics[width=\linewidth]{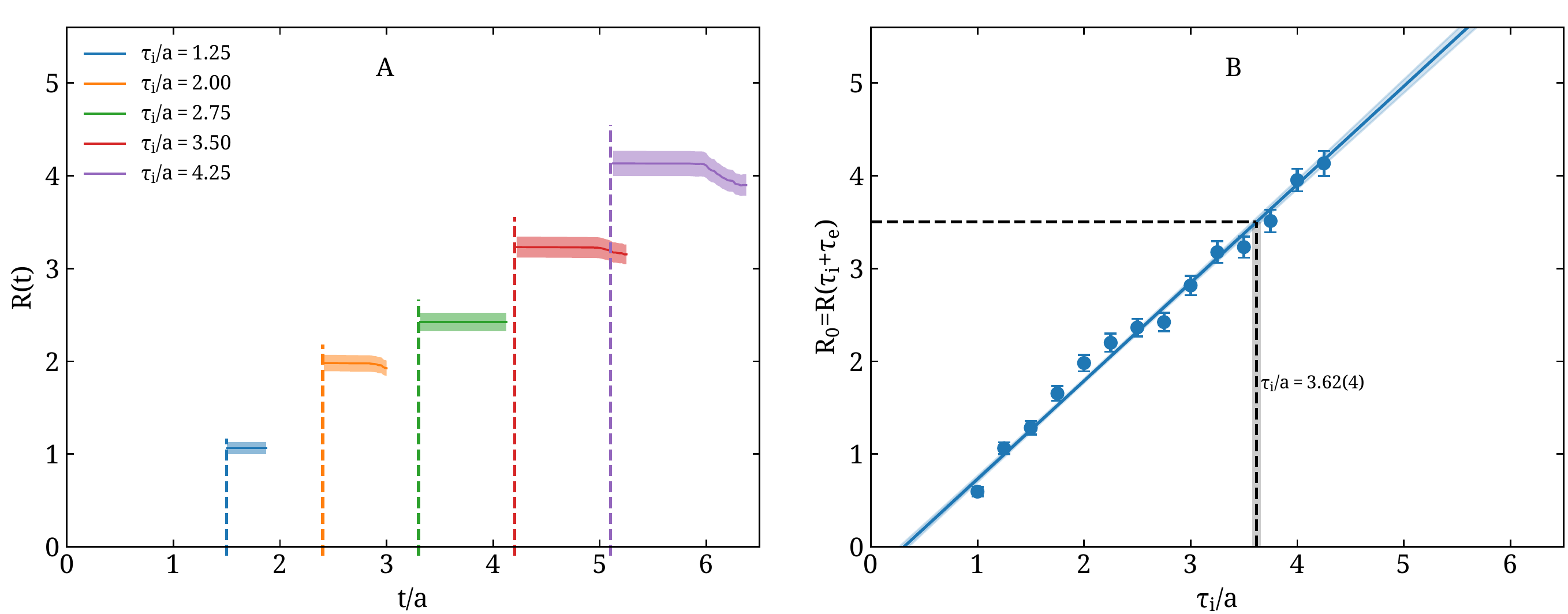}}
  \caption{\textbf{Setting the particle model time scale.} Left panel (A):
    The average number of transmissions per individual as a function
    of time $R(t)$ for representative values of the infectious period
    in simulation time units ($\tau_i/a$), namely for $\tau_i/a=$1.25
    (blue), 2 (orange), 2.75 (green), 3.5 (red), and 4.25
    (purple). The dashed vertical line shows $t=\tau_i+\tau_e$ for
    each case. The errorband in $R(t)$ is from 256 statistics. Right
    panel (B): The basic reproduction number $R_0$ obtained as
    $R(\tau_i+\tau_e)$ from the particle model as a function of the
    infectious period in simulation time units ($\tau_i/a$). The
    exposed time $\tau_e$ is fixed to $\tau_e = \tau_i/5$. The blue
    line and band are the result of a linear fit to $R_0$. The
    horizontal dashed line shows $R_0=3.5$, used to set the scale $a$,
    via $\tau_i = 10$~days.}
  \label{fig:particle scale setting}
\end{figure}

The particle simulation time units are in a scale that can only be
expressed in physical time units \textit{a posteriori}. To determine
the time scale $a$, we initialize with one randomly chosen exposed
individual and measure the average number of transmissions per
individual ($R$) as a function of time, for multiple values of
$\tau_i/a$. By convention, we only measure this quantity for
individuals that have recovered, which means $R(t)=0$ for
$t<\tau_i+\tau_e$. In the left panel of Fig.~\ref{fig:particle scale
  setting}, we show $R(t)$ measured for representative values of
$\tau_i$. The measurement is repeated 256 times, randomly varying the
initial individual exposed each time, which yields the statistical
error in $R(t)$. $\tau_e$ is kept fixed to one fifth of $\tau_i$. From
this analysis we see that $R(t)$ is constant for a time period after
$t=\tau_i+\tau_e$ long enough to obtain a reliable measurement of
$R_0$ for each choice of $\tau_i$. In the right panel of
Fig.~\ref{fig:particle scale setting} we show the measured $R_0$ as a
function of $\tau_i$ and confirm the linear behavior as expected from
Eq.~(\ref{eq:particle R0}). Demanding that initially $R_0=3.5$ and
$\tau_i$=10~days we fix $a$ via a linear fit, which yields
$a\approx 2.763$~days per simulation time unit.

\begin{figure}[!h]
  \figonoff{
  \begin{center}
  \includegraphics[width=0.65\linewidth]{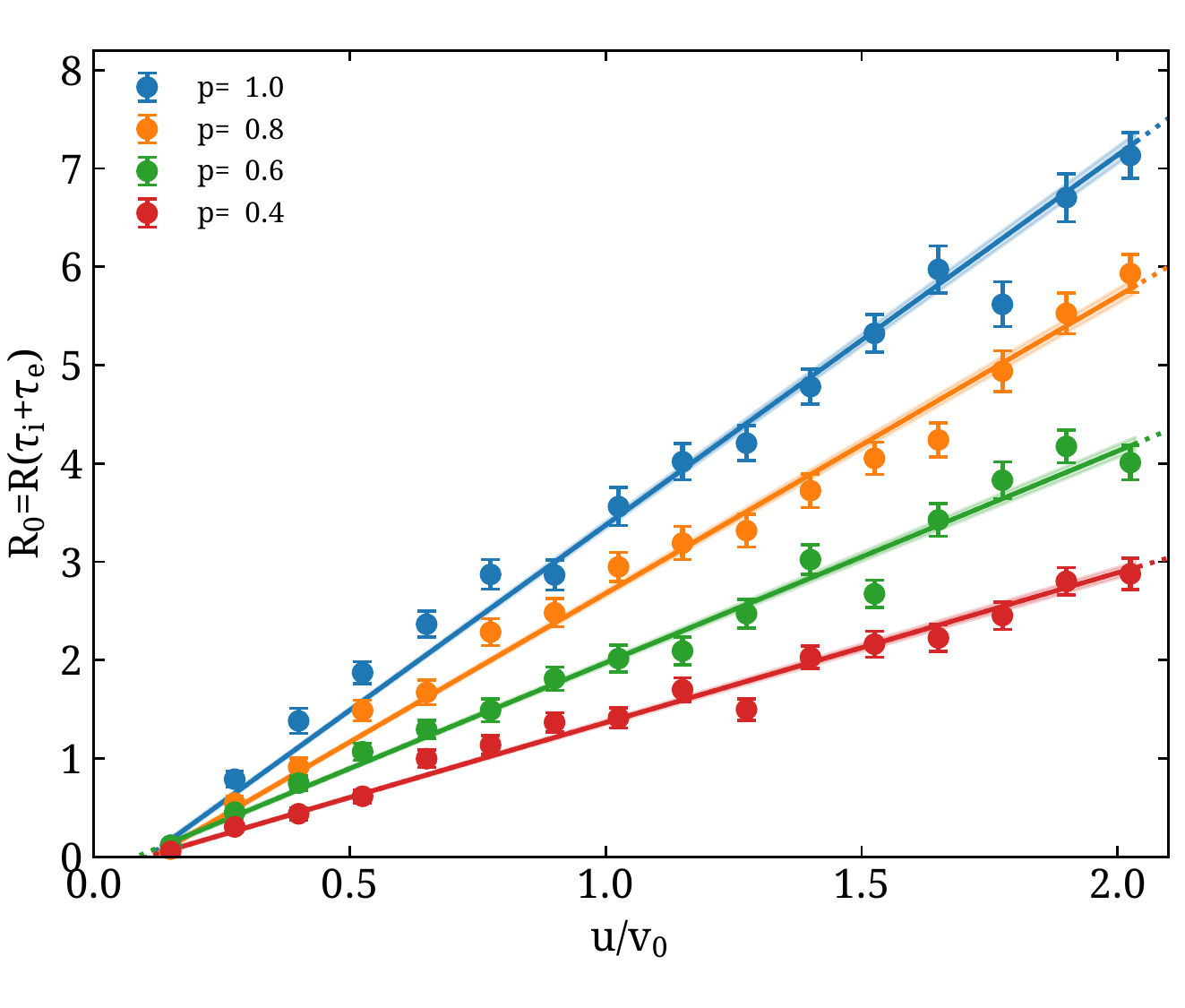}
  \end{center}
  }
  \caption{\textbf{Particle model tuning.} $R_0$ obtained as a function of
    the particle velocities $u/v_0$ for representative values of the
    probability of infection $p$.}
  \label{fig:particle tuning}
\end{figure}

After fixing the time scale $a$ we further tune the particle model in
order to determine the dependence of $R_0$ on the probability of
infection $p$ and velocity scaling factor $u/v_0$. This is shown in
Fig.~\ref{fig:particle tuning}, for representative values of $p$. This
tuning allows us to determine the combination of $u/v_0$ and $p$ to
achieve a desired value of $R_0$. Namely, we use 10 values of $p$
within $[0, 1)$ and fit to the form:
\begin{equation}\label{eq:upfit}
    \alpha_0 \frac{u}{v_0}p + \alpha_1 \frac{u}{v_0} + \alpha_2 p +
    \alpha_3
\end{equation}
obtaining $\vec{\alpha}=(3.68(4), 0.01(2), -0.37(2), -0.012(8))$.

\subsection{Determination and comparison of $R(t)$ between models}
\label{sec:Rt defs}
In the results that follow, we will quote values of the effective
reproduction number $R(t)$ obtained from both the extended SEIQR model
and the particle model. To facilitate comparison, we define the
following:
\begin{itemize}
  \item With $R^\textrm{model}(t)$, for the extended SEIQR model, we
    will quote the value of the effective reproduction number of the
    modeled confirmed cases $C(t)$. This is related to the rate at
    which individuals transition from Susceptible to Quarantined in
    Eq.~(\ref{eq:discrete_system}d), or equivalently the number of new
    confirmed cases per unit time per infective individual
    $(1-r(t))\beta(t)$, multiplied by the period over which the
    individual is infective, $\tau_i$. It is therefore given by
    \begin{equation}
      R^\textrm{model}(t) = \beta(t)(1-r(t))\tau_i,
    \end{equation}
    where $\beta(t)$ is of the form of Eq.~(\ref{eq:beta_t}) and its
    parameters will be determined from fits as will be discussed in
    the next section. For the particle model, we will quote
    $R^\textrm{model}(t)$ as obtained from Eq.~(\ref{eq:upfit}), where
    the velocities $u(t)/v_0$ and probabilities $p(t)$ are
    time-dependent. In particular, for the particle model $R^{\rm
      model}(t)$ is given by
    \begin{equation}
      R^\textrm{model}(t) = \alpha_0 \frac{u(t)}{v_0}p(t) +
      \alpha_1 \frac{u(t)}{v_0} + \alpha_2 p(t) + \alpha_3,
    \end{equation}
    with $\vec{\alpha}$ as determined in Sec.~\ref{sec:particle}.
  \item With $R^\textrm{integral}(t)$ we will quote, for both models,
    an integral definition of $R(t)$ as defined in
    Ref.~\cite{Fodor2020}. Namely, in this case we take
    \begin{equation}
      R^\textrm{integral}(t) = \tau_i\frac{\rho(t+1)}{\sum_{i=t-(\tau_i+\tau_e)+1}^{t-\tau_e}\rho(i)}
      \label{eq:R0 integral}
    \end{equation}
    where $\rho(t)$ are the modeled daily new cases at time $t$ as
    determined by one of our models. With
    $R^{\textrm{integral}}_{(data)}(t)$ we indicate this definition
    used on the actual reported daily new cases of Cyprus, i.e. using
    $\rho(t) = c_t-c_{t-1}$ in Eq.~(\ref{eq:R0 integral}). This
    definition is used since it was shown to account better for the
    fluctuations shown in the reported cases.
\end{itemize}

\section{Results}\label{sec:results}
The two aforementioned models can now be implemented in a
country-specific case. In this study we consider the case of Cyprus
for which such modeling and forecasting is lacking. Our analysis
strategy is to adjust the parameters of the two models to reproduce
the available data of Cyprus up to the writing of this manuscript,
namely July 31\textsuperscript{st}, 2020 and then predict the
evolution until the end of 2020 under different scenarios that reflect
the measures and behaviors known at the time of this writing.  We then
present in an addendum, added after review, the changed situation
where incoming population caused an increased in the positive cases
and new measures. As will be discussed, both models can be adjusted to
take into account the additional measures and predict correctly the
evolution under the updated circumstances.  In what follows we will
model the initial lock-down phase, up until July
31\textsuperscript{st}, 2020, with four different forecasting
scenarios as explained below. The software for generating the plots
that will be presented in this section is available at
Ref.~\cite{swrepo}, while the data are available at
Ref.~\cite{darepo}. Furthermore, the SEIQR model presented here has
been adapted to fit data in real-time and is maintained on an online
platform~\cite{platform}.

\begin{figure}[!h]
  \figonoff{
    \begin{center}
      \includegraphics[width=0.65\linewidth]{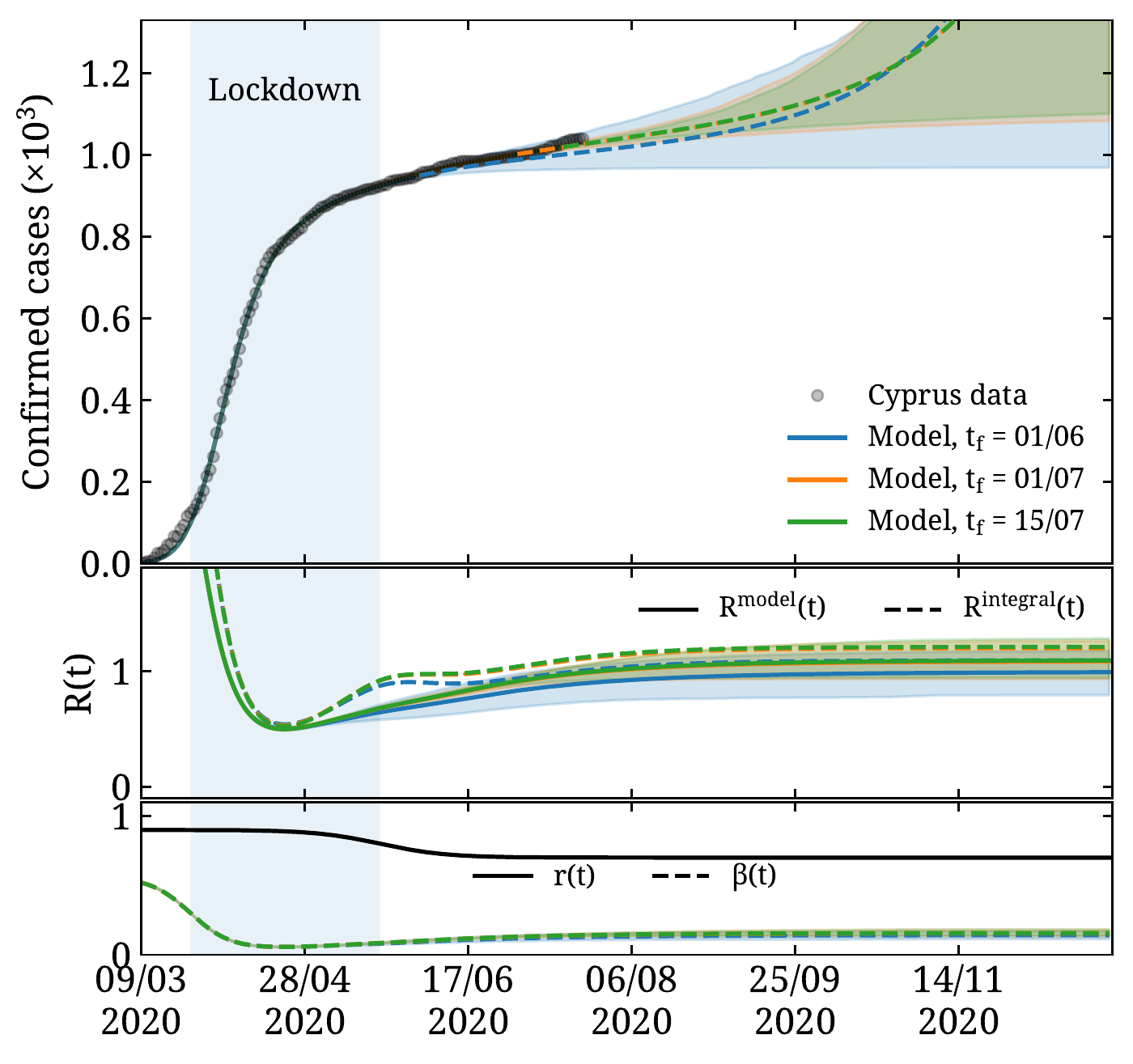}
    \end{center}
  }
  \caption{\textbf{Fitting the SEIQR model.}  In the top panel, we show
    the reported cases in Cyprus (circles) as a function of time and
    the result of fitting our SEIQR model. We consider three periods
    taking $t_\textrm{f}=84$ (blue curve), 114 (orange curve), and 128
    (green curve) that overlap with the data. The blue shaded region
    shows the interval between March 25$^{\rm th}$ and May 21$^{\rm
      st}$, used as $t_1$ and $t_2$ in Eq.~(\ref{eq:beta_t}). The
    dashed curves and bands show the predicted mean value and 90\%
    confidence level as the time range being fitted varies by changing
    $t_\textrm{f}$. The central panel shows $R^\textrm{model}$ with
    the solid curves and $R^\textrm{integral}$ with the dashed
    curves. The bottom panel shows the ratio of undetected $r(t)$.}
  \label{fig:ode t_fit}
\end{figure}

In Fig.~\ref{fig:ode t_fit} we show the result of fitting
Eq.~(\ref{eq:confirmed_t}) of our model to reported COVID-19 positive
cases in Cyprus~\cite{cygov} to obtain the parameters of $\beta(t)$
for the SEIQR model. We validate the model by using data up to date
$t_\textrm{f}$ and compare with the known cases. We take
$t_\textrm{f}$ to be June 1$^{\rm st}$, July 1$^{\rm st}$, and July
15$^{\rm th}$.  We sample the fit parameters via Markov Chain Monte
Carlo using the Stan library~\cite{JSSv076i01}, defined with a
likelihood function via $\e^{-\chi^2/2}$, where $\chi^2$ is given by
\begin{equation}
  \chi^2 = \sum_{t=0}^{t_\textrm{f}} \left|\frac{c_t - C(\vec{\theta}, t)}{\sigma}\right|^2,
  \label{eq:chi2}
\end{equation}
with $c_t$ the cumulative reported cases for Cyprus with $t=0$ March
9\textsuperscript{th} and $C(\vec{\theta}, t)$ the result from the
SEIQR model, namely Eqs.~(\ref{eq:confirmed_t}) or~(\ref{eq:two pop}).
The vector $\vec{\theta}$ contains the parameters to be fitted, namely
$\vec{\theta}=(b_0, b_1, b_2, m_1,m_2)$, with $M=2$ in
Eq.~(\ref{eq:beta_t}) and $\sigma$ drawn from a normal
distribution. We generate 400 independent samples of the parameters
obtained using two chains of 2000 iterations, discarding the first
1000 as thermalization iterations and of the remaining 1000 we took
every fifth in each chain.

At this point in time, the number of actual cases as compared to the
reported are not known.  For example, an early study analyzing testing
data from Iceland put this number between five to ten times. There,
intensive testing was carried out early on, obtained from
1\textsuperscript{st} to 4\textsuperscript{th} of
April~\cite{Stock2020.04.06.20055582}. A study in the US, analyzing
data obtained from 23\textsuperscript{rd} March to
12\textsuperscript{th} May, estimated this number to be between six to
24 times~\cite{10.1001/jamainternmed.2020.4130}.

\begin{figure}[!h]
  \figonoff{
    \begin{center}
      \includegraphics[width=0.65\linewidth]{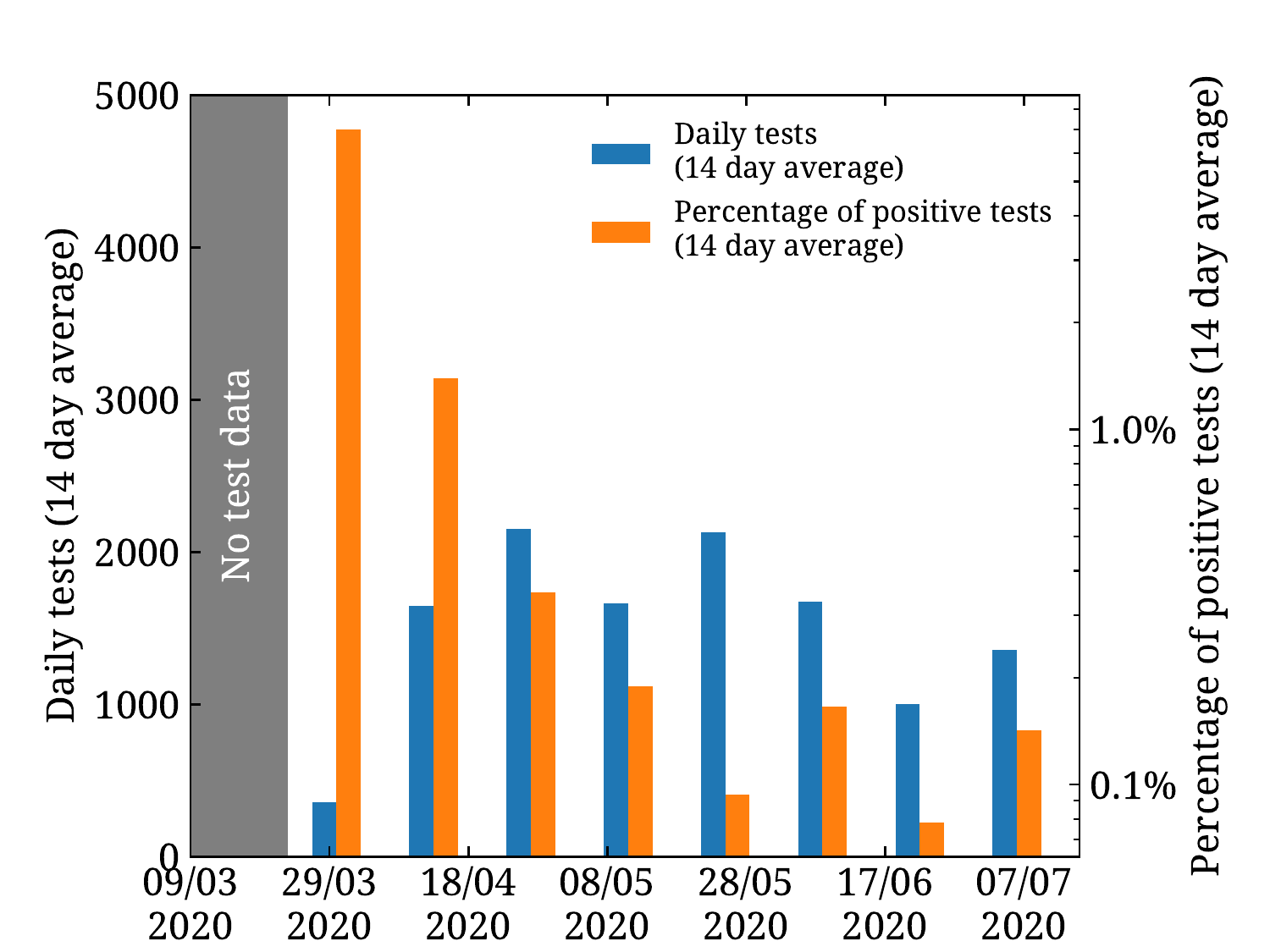}
    \end{center}
  }
  \caption{\textbf{Daily tests and positivity rate in Cyprus.} 14-day
    average of the daily tests (blue bars) and of the percentage of
    daily tests returned positive (orange bars) for Cyprus. The former
    is plotted on the left vertical axis while the latter on the right
    vertical axis.}
  \label{fig:tests}
\end{figure}

In Cyprus, we observe an intensification of testing and tracing since
May~\cite{cygov} and a low death rate~\cite{owidcoronavirus}, which
indicates low prevalence of the virus. We take the portion of
undetected cases $r(t)$, to gradually decrease from one detected in 10
infected to one in three by using the functional form for the portion
of undetected cases $r(t)$ to be
\begin{equation}
  r(t) = \frac{r_1+r_0}{2} + \frac{r_1-r_0}{2}\tanh(m_r(t-t_r)),
\end{equation}
with $r_0=0.9$, $r_1=0.7$, $m_r=0.05$, and $t_r=73$ or May
21\textsuperscript{st}, 2020. This choice is motivated by the
evolution of daily tests and positivity rate, shown in
Fig.~\ref{fig:tests}, which stabilize around May to about 2000 tests
per day at a positivity rate of about 0.1\%~\cite{cydata}.

As can be seen, we obtain consistent results and good prediction of
the available data when fitting until June 1\textsuperscript{st}
i.e. when omitting up to two months of the most recent data from the
fits. From the central panel, we see that the estimates for $R(t)$
overlap when varying $t_\textrm{f}$, which confirms that our fits
yield consistent parameters for the choices considered. The parameters
obtained for each fit range are tabulated in Table~\ref{tab:ode
  parameters}.

\begin{table}[H]
  \caption{Fit parameters obtained for $\beta(t)$ for three choices of
    the final day $t_\textrm{f}$ used in the fit, namely for day 84
    (June 1\textsuperscript{st}), 114 (July 1\textsuperscript{st}),
    and 128 (July 15\textsuperscript{th}), with $t=0$ being April
    9\textsuperscript{th}. Errors are quoted with the superscript and
    subscript for the upper and lower bounds respectively, obtained
    via a Markov chain Monte Carlo as explained in the
    text.}\label{tab:ode parameters}
  \begin{tabular}{rccccc}
    \hline\hline
    $t_\textrm{f}$ [days] & $b_0$ [days$^{-1}$] & $b_1$ [days$^{-1}$] & $b_2$ [days$^{-1}$] & $m_1$ [days$^{-1}$] & $m_2$   [days$^{-1}$]\\
    \hline
    84  & 0.550$^{+0.006}_{-0.001}$   & 0.019$^{+0.009}_{-0.005}$    & 0.143$^{+0.009}_{-0.009}$   & 0.082$^{+0.001}_{-0.002}$    & 0.019$^{+0.002}_{-0.004}$ \\
    114 & 0.553$^{+0.004}_{-0.001}$   & 0.015$^{+0.011}_{-0.005}$    & 0.155$^{+0.005}_{-0.011}$   & 0.080$^{+0.001}_{-0.001}$    & 0.020$^{+0.002}_{-0.003}$ \\
    128 & 0.553$^{+0.005}_{-0.001}$   & 0.014$^{+0.012}_{-0.005}$    & 0.157$^{+0.005}_{-0.001}$   & 0.080$^{+0.001}_{-0.001}$    & 0.020$^{+0.002}_{-0.004}$ \\
    \hline\hline
  \end{tabular}
\end{table}

\begin{figure}[!h]
  \figonoff{
    \begin{center}
      \includegraphics[width=0.65\linewidth]{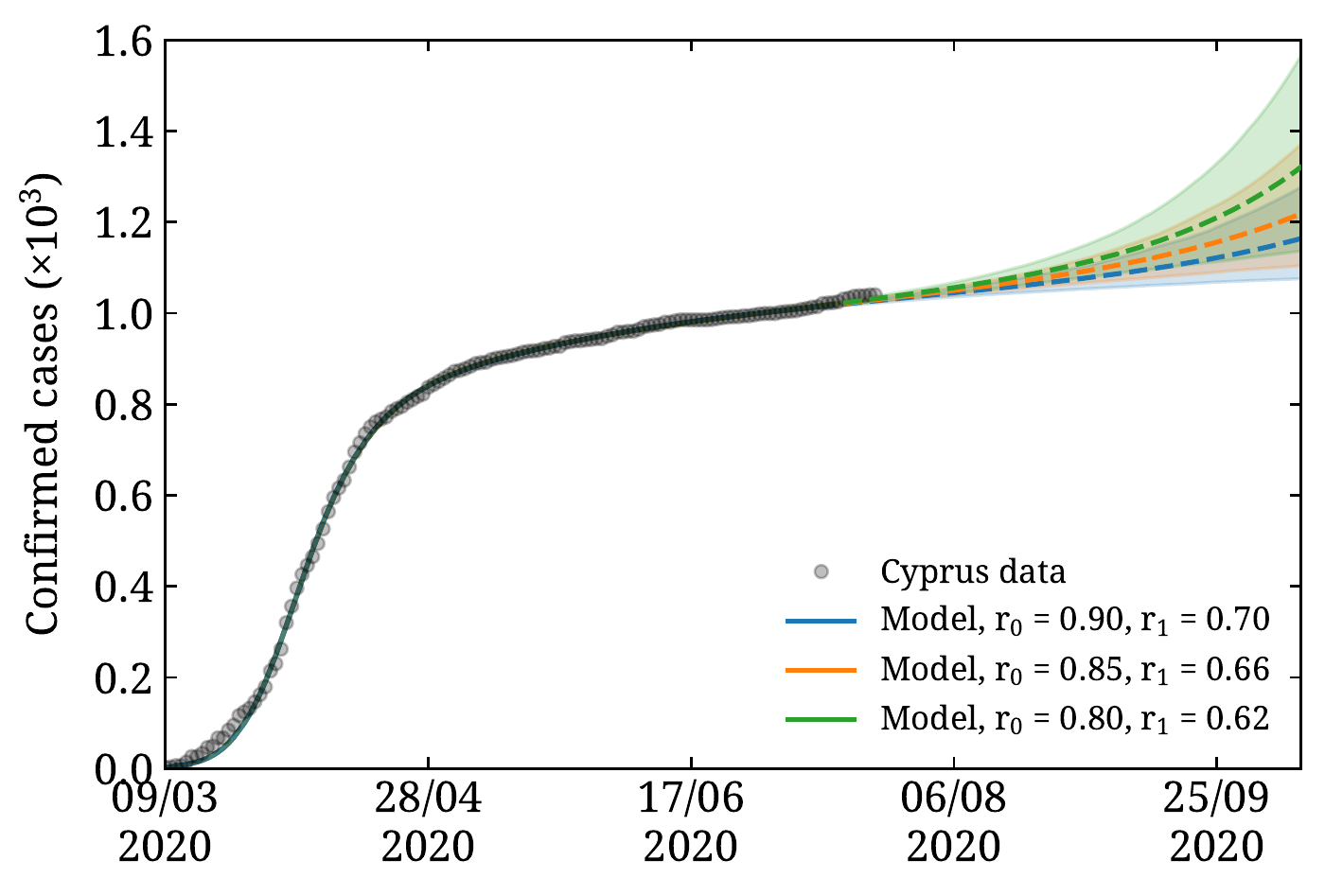}
    \end{center}
  }
  \caption{\textbf{Sensitivity to detection rate.}  The result of fitting
    our SEIQR model using $(r_0, r_1)$ = (0.9, 0.7) (blue curve and
    band), (0.85, 0.66) (orange curve and band), and (0.80, 0.62)
    (green curve and band). In all cases the fit includes confirmed
    cases (black points) until $t_\textrm{f}=$128 (July
    15\textsuperscript{th}).}
  \label{fig:r_cmp}
\end{figure}

As mentioned, we use $r_0=0.9$ and $r_1=0.7$ motivated by the ramping
up of daily testing in Cyprus during May 2020. We note that our fits
are robust to small changes to $r_0$ and $r_1$ as long as their ratio
is maintained and excluding extrema such as $r_0\simeq 1$ and
$r_1\simeq 0$. In Fig.~\ref{fig:r_cmp} we compare two choices for
$r_0$ and $r_1$ in addition to those used in Fig.~\ref{fig:ode t_fit}
for the case of $t_\textrm{f}=$128. We see that indeed the evolution
of the pandemic up to the end of October 2020 is within errors for all
three choices, and thus that our predictions are not sensitive to the
precise inputs for $r_0$ and $r_1$. For the remainder of this paper we
will use the choice $r_0=0.9$ and $r_1=0.7$.

Having validated our extended SEIQR model by predicting the reported
cases over the known period by varying $t_\textrm{f}$, we make
forecasting using four plausible scenarios:
\begin{description}
\item [A)] In scenario A, we forecast the evolution of the pandemic
  assuming infection and testing rates remain unchanged, i.e. we take
  $\beta(t)$ and $r(t)$ to remain the same after the fitted period.
\item [B)] In scenario B, we investigate the case where $r(t)$
  asymptotically reaches 50\% by mid August. This reflects a scenario in
  which more aggressive testing and contact tracing is performed.
\item [C)] In scenario C, we decrease $\beta(t)$ such that
  $R^\textrm{model}(t)$ approximately reaches 1 by mid August. Thus we
  model the case in which measures are tightened in order to preempt a
  second wave.
\item [D)] In scenario D, we use Eq.~(\ref{eq:two pop}) to model the
  effect of splitting the population into two groups. Namely, for 80\%
  of the population we take $\beta(t)$ to remain unchanged as in
  scenario~A, with the remaining 20\% having a $\beta(t)$ that
  reproduces $R^\textrm{model}(t)\simeq 0.5$. This scenario models a
  situation in which, for example, restrictions are imposed on
  vulnerable groups.
\end{description}

These four scenarios are useful in demonstrating how we can modify
features of our models to gain insights on the evolution of the
pandemic, however they are idealized in that they assume smooth
changes of the infection rate over long periods of time. Long-term
predictions about the evolution of the disease largely depend on
factors the modeling of which is beyond the scope of this work, such
as policy and public response to imposed measures, availability of
vaccinations, and travel. In the addendum of this paper we address how
longer time series of available data can be modeled for short-term
predictions, using data that have become available while this
manuscript was under review.

\begin{figure}[H]
  \figonoff{
    \includegraphics[width=\linewidth]{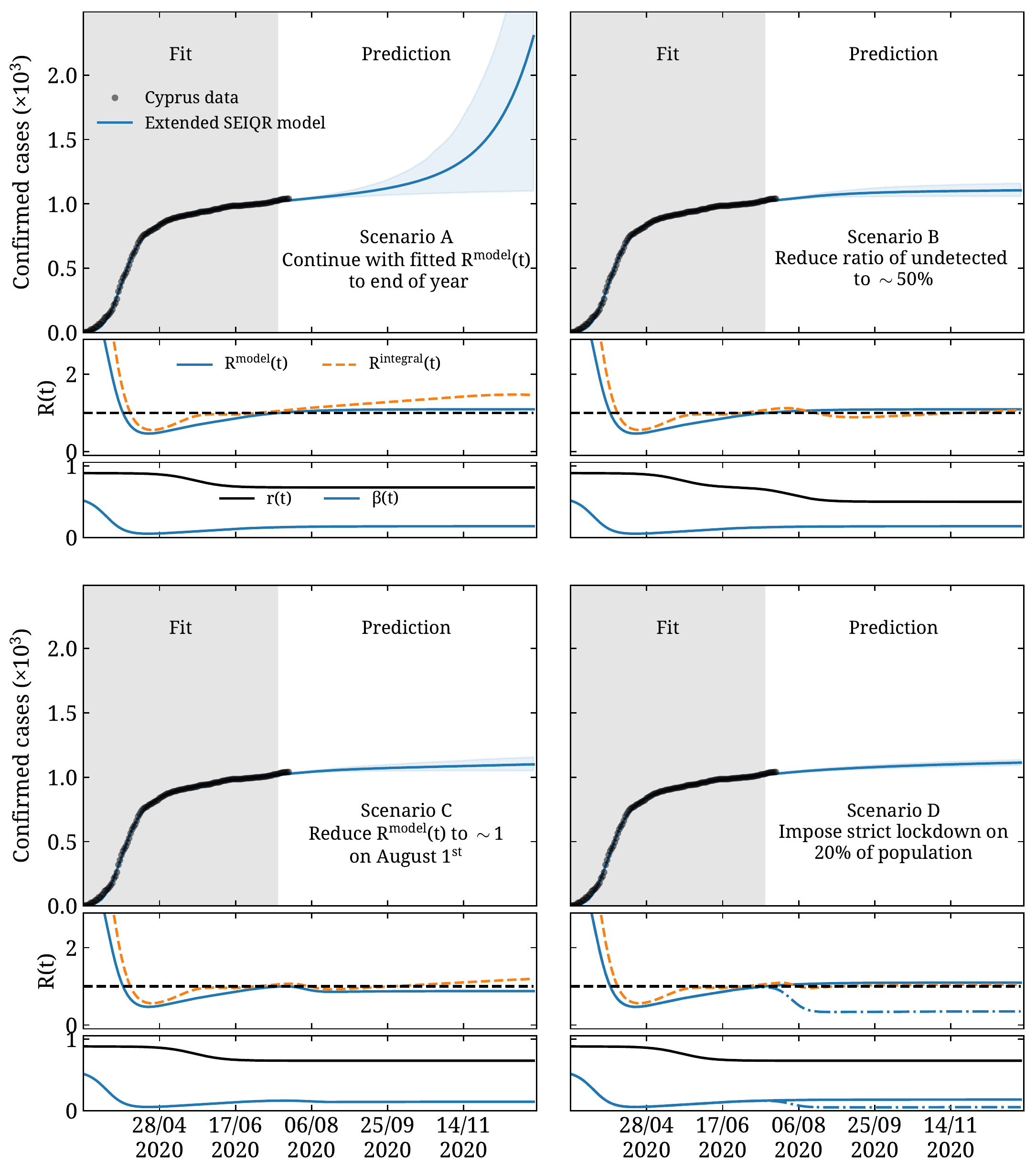}
  }
  \caption{\textbf{Forecasting using the SEIQR model.} Predicting the
    evolution of COVID-19 for scenarios A and B (upper panel) and C
    and D (lower panel) obtained using our extended SEIQR model. In
    the upper part of each plot we show the reported cases in Cyprus
    (circles) as a function of time.  The gray band shows the period
    used to fit the parameters, namely we use data on the reported
    cases up to $t_\textrm{f}$=128 or July 15\textsuperscript{th}. The
    blue curve and band shows our prediction for each scenario. In the
    central part of each plot, we show with the solid blue curve
    $R^\textrm{model}(t)$, and with the dashed yellow curve
    $R^\textrm{integral}(t)$. For the case of scenario D, the solid
    blue curve corresponds to $R^\textrm{model}(t)$ evaluated using
    $\beta(t)$ used for 80\% of the population, while the dash-dotted
    line shows $R^\textrm{model}(t)$ evaluated using $\beta(t)$ used
    for 20\%. In the bottom part we show $r(t)$ (black curve) and
    $\beta(t)$ (blue curve).  }
  \label{fig:ode scenarios}
\end{figure}

The forecasting for the four scenarios using the SEIQR model are shown
in Fig.~\ref{fig:ode scenarios}, when using the parameters as
determined by fitting the Cyprus reported cases taking
$t_\textrm{f}$=128 or July 15$^{\rm th}$. As can be seen, for scenario
A, the mean value of infected steadily increases. In scenarios B, C,
and D on the other hand, within the uncertainty bands, we observe a
flattening of the daily cases until end of the year.

\begin{figure}[H]
  \figonoff{
    \includegraphics[width=\linewidth]{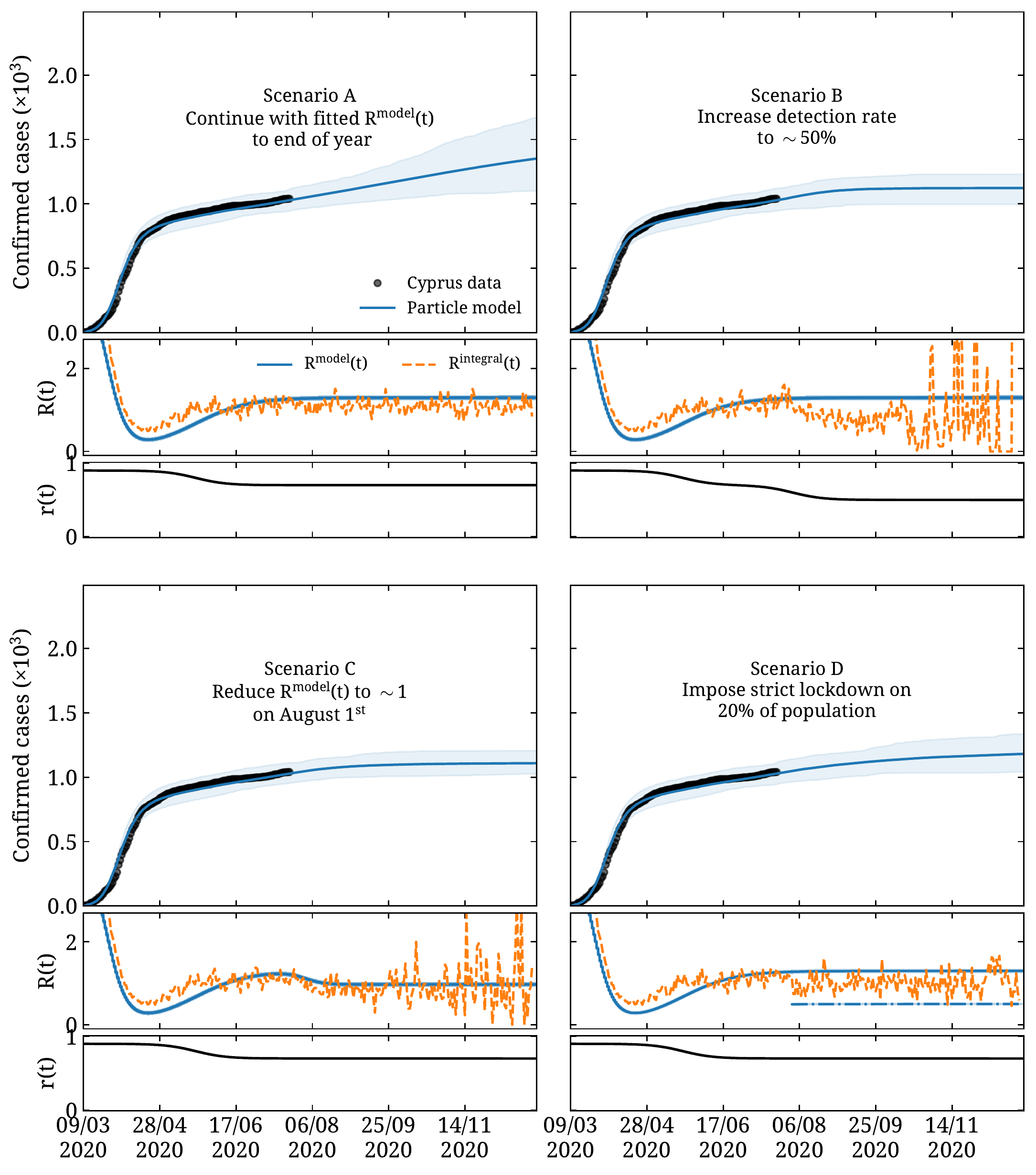}
  }
  \caption{{\bf Forecasting using the particle model.} Predictions for
    the same four scenarios as in Fig.~\ref{fig:ode scenarios} but
    using the particle model. In the top part of the plots, the blue
    curve and associated band is obtained as the average and 90\%
    confidence interval when seeding the model 32 times. In the
    central part of the plots, we show with the blue curve
    $R^\textrm{model}(t)$ and with the orange dashed curve
    $R^\textrm{integral}(t)$. The rest of the notation is the same as
    that used in Fig.~\ref{fig:ode scenarios}.}
  \label{fig:particle scenarios}
\end{figure}

Our predictions based on the time evolution of the particle model for
the four different scenarios are shown in Fig.~\ref{fig:particle
  scenarios}. We use the same $r(t)$ as used in the extended SEIQR
model. For the functional form of $u(t)$, we use $\beta(t)$, where now
instead of $b_j$ of Eq.~(\ref{eq:beta_t}) we adjust the particle
velocities $u_j$ for $j=0,$ 1, and 2 to minimize $\chi^2$ defined in
the same way as in the SEIQR model given in Eq.~(\ref{eq:chi2}). The
error bands in the upper part of the plots are statistical, obtained
as the 90\% confidence level from independently seeding the particle
model 32 times. Requiring an increase of the minimum $\chi^2$ yields
errors in the parameters that are smaller than 0.5\% and are therefore
negligible compared to the statistical errors shown in
Fig.~\ref{fig:particle scenarios}.

As can be seen from comparing the forecasts shown in
Figs.~\ref{fig:ode scenarios} and~\ref{fig:particle scenarios}, the
two models are qualitatively in agreement in their predictions for all
four scenarios. In particular, they both predict that if infection
rates remain the same as on July 15\textsuperscript{th}, within
scenario A and within the uncertainty one cannot exclude either a
steady future increase in daily cases that can turn into an epidemic
or a flattening. Namely, for the compartmental model, the number of
predicted cumulative cases for the 31\textsuperscript{st} of Dec. has
central value 2302 with 90\% confidence interval yielding the range
1101 to 3556. For the particle model, the same forecast is 1352
with range 1099 to 1674.

For the other three scenarios, which assume measures are taken that
will decrease infection rate and/or increase testing, both models show
a flattening or complete suppression of daily cases by the end of the
year. It should be noted that the large fluctuations observed in
$R^\textrm{integral}(t)$ in Fig.~\ref{fig:particle scenarios} arise
from the discrete nature of the particle model when multiple
consecutive days yield zero new cases.

While this work was under review Cyprus experienced a second wave,
resulting in confirmed cases rising at a larger rate than our worst
case, scenario~A. In the addendum we address how we model the longer
term time evolution of the pandemic and compare to scenario~A.

\section{Conclusions}\label{sec:conclusions}
In this work we developed two different approaches to model the
evolution of COVID-19. The two approaches can be used in any country
and do not required large numbers of data. In particular they can be
applied in countries like Cyprus, where the most consistent and
reliable data are for the daily reported cases since data on
e.g. deaths, intubations, and hospitalizations are too few for a
meaningful statistical analysis. The models are highly complementary;
the SEIQR model is an ODE-based compartmental approach and is
computationally fast allowing sampling of parameters over long Monte
Carlo Markov chains; the particle simulation approach, while
computationally demanding, allows for tagging and tracing individuals
and changing infection probabilities for arbitrary subsets of the
population.

We calibrate and validate the models using the Cyprus reported
positive COVID-19 data. For both models, we are able to fit the daily
reported cases using five parameters that describe either the change
of infection rate over time for the case of the SEIQR model or the
velocities of the particles for the case of the particle model. In
addition, we are able to obtain consistent results when using the two
models to predict the evolution of COVID-19 for four scenarios that can
be applied to control the epidemic.

The four scenarios are chosen as appropriate examples to demonstrate
the range of parameters that can be adjusted using our two models,
such as future lockdowns or easing of restrictions for subsets of the
population and changes in the number of tests performed that in turn
change the ratio of infected reported. Richer scenarios can be
forecasted by combining these measures and by selectively applying
them to multiple subsets of the population.

Although the robustness of the model fits we have undertaken pertains
to datasets from other countries, we have chosen to limit the
discussion to the case for Cyprus for the sake of
brevity. Furthermore, future work to enhance the models will include
allowing for non-uniform spatial distributions to model different
population densities and data-driven modeling of the detection rate as
more data become available.

\section*{Addendum}
\begin{figure}[H]
  \figonoff{
    \includegraphics[width=0.5\linewidth]{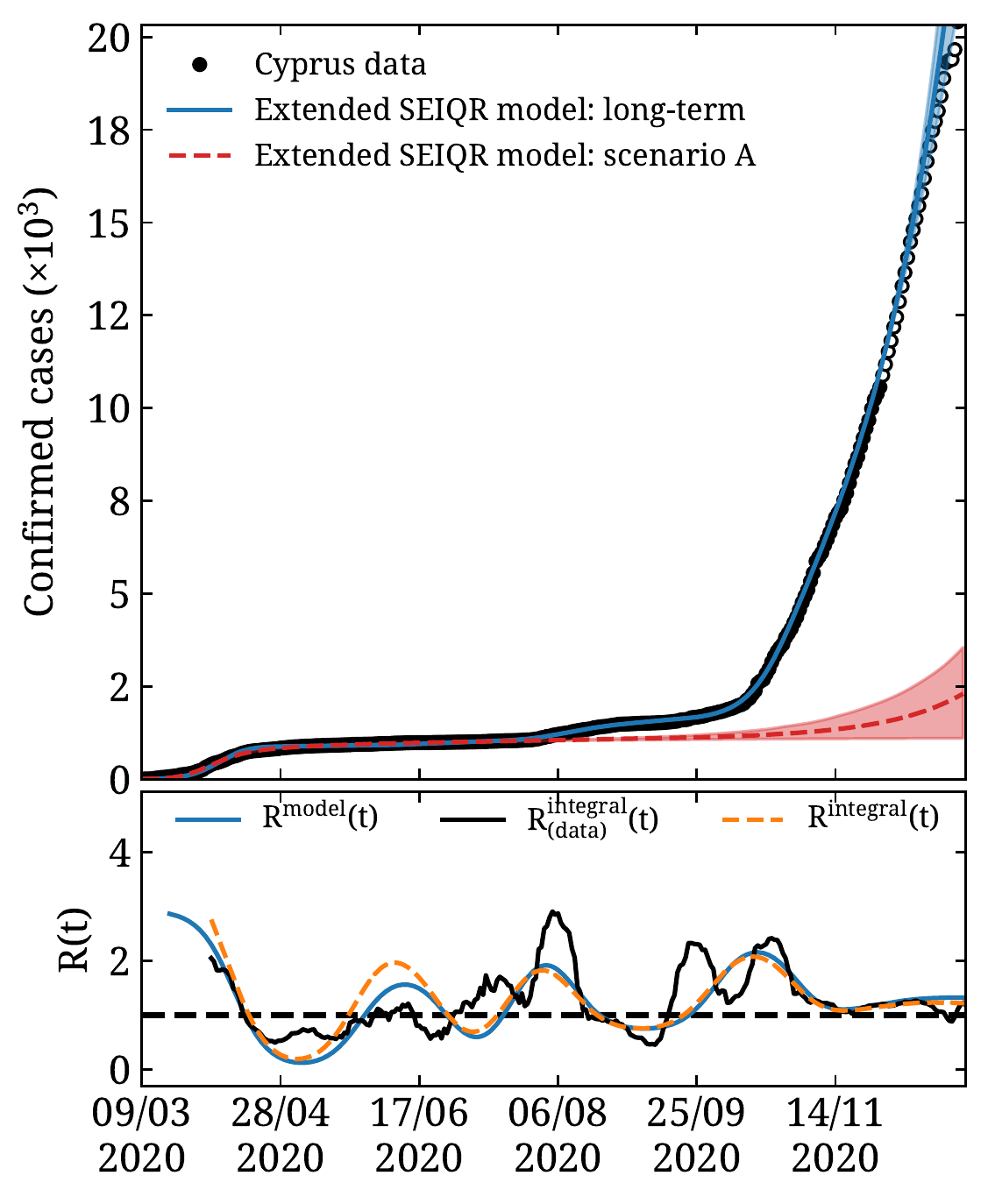}
    \includegraphics[width=0.5\linewidth]{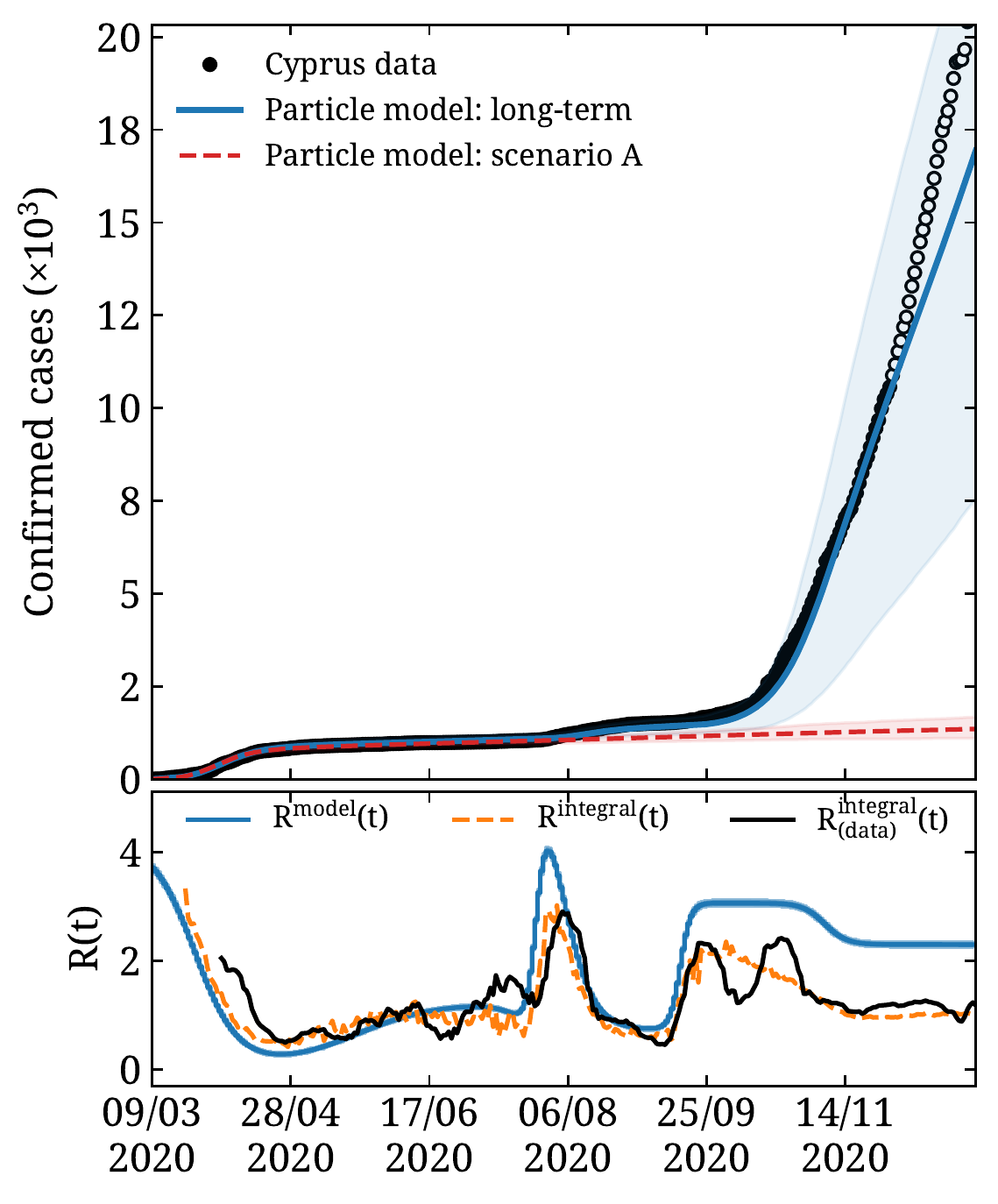}    
  }
  \caption{{\bf Modeling of longer time series.} The SEIQR model
    (left) and the particle model (right) fitted to the Cyprus
    cumulative confirmed cases up to November 30\textsuperscript{th}
    (solid blue curve and band) as explained in the addendum and
    compared to scenario~A (dashed red curve and band). The filled
    black circles are the Cyprus confirmed cases included in the fit,
    while the open black circles are not included in the fit. The
    forecast is for one month, {\it assuming the infection rate stays
      constant}. In the lower panels we show the effective
    reproduction number as obtained by applying Eq.~(\ref{eq:R0
      integral}) to the Cyprus confirmed cases denoted as
    $R^\textrm{integral}_{\textrm{(data)}}(t)$ (black curve), in
    addition to $R^\textrm{model}(t)$ (blue curve) and
    $R^\textrm{integral}(t)$ (orange dashed curve).}
  \label{fig:models_long}
\end{figure}

As mentioned in Sec.~\ref{sec:models}, for longer time series data we
determine the number of inflection points $M$ and their values $t_j$
by analyzing the second derivative of the confirmed cases $c_t$. To
demonstrate this approach, we carry out an analysis fitting the SEIQR
model to data until November 30\textsuperscript{th} to obtain $b_j$,
$j=0, ..., M$ with $M=7$ and $m_j=0.08$ fixed. We note that these
inflection points are obtained by analyzing the confirmed cases
reported and the interpretation of their causes is non-trivial and
beyond the scope of our modeling, given that they depend not only on
the imposition and lifting of measures but also to factors such as the
response of society to the measures. The result is shown in the left
panel of Fig.~\ref{fig:models_long}, where we see that the existing
data are described well by the model. Comparison with scenario~A,
which assumed a long-term, constant infection rate, reveals a
worsening of the situation compared to the assumptions made within
that scenario. In the lower panel, we show $R^\textrm{model}(t)$ and
$R^\textrm{integral}(t)$, which are obtained from the modeled
confirmed cases as in Figs.~\ref{fig:ode scenarios}
and~\ref{fig:particle scenarios}. We also show
$R^\textrm{integral}_{\textrm{(data)}}(t)$, obtained by using a
rolling 14 day average of the Cyprus data for the daily confirmed
cases $\rho(t)$ in Eq.~(\ref{eq:R0 integral}). We compare these three
definitions of $R(t)$ to asses whether the number of inflection points
identified is consistent with the underlying data. As can be seen, all
three definitions are qualitatively in agreement displaying similar
main features. For the particle model, shown in the right panel of
Fig.~\ref{fig:models_long}, we similarly adjust the velocities to
reproduce the daily confirmed cases $c_t$. The same comparison is made
between the three definitions of $R(t)$ that are again in qualitative
agreement. In both cases, the one month forecast made {\it assuming
  the infection rate remains constant}, i.e. similar to scenario~A
above, yields consistent predictions for the number of confirmed cases
until the end of 2020 within the two model uncertainties. This
demonstrates how the models can capture the behavior of longer time
series, by adjusting the parameters of both models. We are able to
reproduce the data with a minimal set of parameters and both models
agree in the forecasted one month trajectory of the pandemic assuming
infection rates do not change.

\section*{Acknowledgments}
We would like to thank Maria Koliou Mazeri for fruitful communication
and exchanges during the authoring of this paper. AI acknowledges
support from the project ``Modeling of the COVID-19 pandemic for
Cyprus'' with contract number CONCEPT-COVID/0420/0011 funded by the
Cyprus Research and Innovation Foundation. VH and NS acknowledge
support by project ``SimEA'', funded by the European Union's Horizon
2020 research and innovation programme under grant agreement No
810660.

\nolinenumbers

%% \bibliography{refs}

\begin{thebibliography}{10}

\bibitem{JHC}
Dong E, Du H, Gardner L.
\newblock An interactive web-based dashboard to track COVID-19 in real time.
\newblock The Lancet Infectious Diseases. 2020;20:533--534.
\newblock doi:{10.1016/S1473-3099(20)30120-1}.

\bibitem{owidcoronavirus}
Max~Roser EOO Hannah~Ritchie, Hasell J.
\newblock Coronavirus Pandemic (COVID-19).
\newblock Our World in Data. 2020;.

\bibitem{Ferguson2020}
Ferguson N, et~al.. Report 9: Impact of non-pharmaceutical interventions (NPIs)
  to reduce COVID19 mortality and healthcare demand; 2020.
\newblock Available from: \url{https://doi.org/10.25561/77482}.

\bibitem{Wang2020}
Wang Y, Chen Y, Qin Q.
\newblock Unique epidemiological and clinical features of the emerging 2019
  novel coronavirus pneumonia (COVID-19) implicate special control measures.
\newblock Journal of Medical Virology. 2020;92:25.

\bibitem{Kupferschmidt2020}
Kupferschmidt K.
\newblock Why do some COVID-19 patients infect many others, whereas most
  don’t spread the virus at all?
\newblock Science. 2020;doi:{10.1126/science.abc8931}.

\bibitem{Giordano2020}
Giordano G, Blanchini F, Bruno R, Colaneri P, Di~Filippo A, Di~Matteo A, et~al.
\newblock Mod- elling the COVID-19 epidemic and implementation of
  population-wide interventions in Italy.
\newblock Nature Medicine. 2020;doi:{10.2038/s41591-020-0883-7}.

\bibitem{Gatto2020}
Gatto M, Bertuzzo E, Mari S L amd~Miccoli, Carraro L, Casagrandi R, Rinaldo A.
\newblock Effects of emergency contain- ment measures.
\newblock Proceedings of the National Academy of Sciences. 2020;117:10484.

\bibitem{Ferguson2020b}
Seth F, et~al.. Report 13: Estimating the number of infections and the impact
  of non-pharmaceutical interventions on COVID-19 in 11 European countries;
  2020.
\newblock Available from: \url{doi: https://doi.org/10.25561/77731}.

\bibitem{PPR:PPR113604}
Peng L, Yang W, Zhang D, Zhuge C, Hong L. Epidemic analysis of COVID-19 in
  China by dynamical modeling; 2020.
\newblock Available from: \url{https://europepmc.org/article/PPR/PPR113604}.

\bibitem{DAS2021110595}
Das A, Dhar A, Goyal S, Kundu A, Pandey S.
\newblock COVID-19: Analytic results for a modified SEIR model and comparison
  of different intervention strategies.
\newblock Chaos, Solitons \& Fractals. 2021;144:110595.
\newblock doi:{https://doi.org/10.1016/j.chaos.2020.110595}.

\bibitem{Keskinocak2020}
Keskinocak P, et~al.. The Impact of Social Distancing on COVID19 Spread: State
  of Georgia Case Study; 2020.
\newblock Available from: \url{https://doi.org/10.1101/2020.04.29.20084764}.

\bibitem{Keskinocak2020b}
Keskinocak P.
\newblock COVID-19 pandemic modeling is fraught with uncertainties.
\newblock Physics Today. 2020;73:25.
\newblock doi:{10.10631/PT.3.4493}.

\bibitem{SIR1927}
Kermack WO, McKendrick AG.
\newblock A contribution to the mathematical theory of epidemics.
\newblock Proceedings of the Royal Society of London Series A.
  1927;115(772):700--721.
\newblock doi:{10.1098/rspa.1927.0118}.

\bibitem{CARLETTI2020100034}
Carletti T, Fanelli D, Piazza F.
\newblock COVID-19: The unreasonable effectiveness of simple models.
\newblock Chaos, Solitons \& Fractals: X. 2020;5:100034.
\newblock doi:{https://doi.org/10.1016/j.csfx.2020.100034}.

\bibitem{FANELLI2020109761}
Fanelli D, Piazza F.
\newblock Analysis and forecast of COVID-19 spreading in China, Italy and
  France.
\newblock Chaos, Solitons \& Fractals. 2020;134:109761.
\newblock doi:{https://doi.org/10.1016/j.chaos.2020.109761}.

\bibitem{Dehningeabb9789}
Dehning J, Zierenberg J, Spitzner FP, Wibral M, Neto JP, Wilczek M, et~al.
\newblock Inferring change points in the spread of COVID-19 reveals the
  effectiveness of interventions.
\newblock Science. 2020;doi:{10.1126/science.abb9789}.

\bibitem{ecdpc}
European Centre for Disease Prevention and Control;.
\newblock Available from:
  \url{https://www.ecdc.europa.eu/en/covid-19/questions-answers}.

\bibitem{Hethcote1980}
Hethcote HW, Tudor DW.
\newblock Integral equation models for endemic infectious diseases.
\newblock Journal of Mathematical Biology. 1980;9(1):37--47.
\newblock doi:{10.1007/bf00276034}.

\bibitem{Wearing2005}
Wearing HJ, Rohani P, Keeling MJ.
\newblock Appropriate Models for the Management of Infectious Diseases.
\newblock {PLoS} Medicine. 2005;2(7):e174.
\newblock doi:{10.1371/journal.pmed.0020174}.

\bibitem{Feng2000}
Feng Z, Thieme HR.
\newblock Endemic Models with Arbitrarily Distributed Periods of Infection {I}:
  Fundamental Properties of the Model.
\newblock {SIAM} Journal on Applied Mathematics. 2000;61(3):803--833.
\newblock doi:{10.1137/s0036139998347834}.

\bibitem{Lloyd2001}
Lloyd AL.
\newblock Realistic Distributions of Infectious Periods in Epidemic Models:
  Changing Patterns of Persistence and Dynamics.
\newblock Theoretical Population Biology. 2001;60(1):59--71.
\newblock doi:{10.1006/tpbi.2001.1525}.

\bibitem{Zhang2008}
Zhang F, zhen Li Z, Zhang F.
\newblock Global stability of an {SIR} epidemic model with constant infectious
  period.
\newblock Applied Mathematics and Computation. 2008;199(1):285--291.
\newblock doi:{10.1016/j.amc.2007.09.053}.

\bibitem{cygov}
Epidemiological Surveillance~Unit MoH. Cyprus National Situation Report; 2020.
\newblock Available from: \url{https://www.pio.gov.cy/coronavirus/en/}.

\bibitem{doi:10.1002/jcc.21915}
Bannerman MN, Sargant R, Lue L.
\newblock DynamO: a free \${\cal O}\$(N) general event-driven molecular
  dynamics simulator.
\newblock Journal of Computational Chemistry. 2011;32(15):3329--3338.
\newblock doi:{10.1002/jcc.21915}.

\bibitem{pSEIQR}
p-SEIQR;.
\newblock Available from: \url{https://github.com/g-koutsou/p-SEIQR}.

\bibitem{Fodor2020}
Fodor Z, Katz SD, Kovacs TG. Why integral equations should be used instead of
  differential equations to describe the dynamics of epidemics;.

\bibitem{swrepo}
Software repository;.
\newblock Available from: \url{https://github.com/g-koutsou/2008.03165}.

\bibitem{darepo}
Data repository;.
\newblock Available from: \url{https://osf.io/bp79h/}.

\bibitem{platform}
COVID-19 Modeling Platform;.
\newblock Available from: \url{http://covid-model-platform.cyi.ac.cy}.

\bibitem{JSSv076i01}
Carpenter B, Gelman A, Hoffman M, Lee D, Goodrich B, Betancourt M, et~al.
\newblock Stan: A Probabilistic Programming Language.
\newblock Journal of Statistical Software, Articles. 2017;76(1):1--32.
\newblock doi:{10.18637/jss.v076.i01}.

\bibitem{Stock2020.04.06.20055582}
Stock JH, Aspelund KM, Droste M, Walker CD.
\newblock Identification and Estimation of Undetected COVID-19 Cases Using
  Testing Data from Iceland.
\newblock medRxiv. 2020;doi:{10.1101/2020.04.06.20055582}.

\bibitem{10.1001/jamainternmed.2020.4130}
Havers FP, Reed C, Lim T, Montgomery JM, Klena JD, Hall AJ, et~al.
\newblock {Seroprevalence of Antibodies to SARS-CoV-2 in 10 Sites in the United
  States, March 23-May 12, 2020}.
\newblock JAMA Internal Medicine. 2020;doi:{10.1001/jamainternmed.2020.4130}.

\bibitem{cydata}
Daily Statistics of the Evolution of COVID-19 in Cyprus;.
\newblock Available from: \url{https://www.data.gov.cy/node/4617?language=en}.

\end{thebibliography}

\end{document}